\documentclass{elsart3p}

\usepackage{natbib}

\usepackage{graphicx}

\usepackage{amssymb}

\def\aap{A\&A }

\def\apj{ApJ }
\def\apjl{ApJ }
\def\mnras{MNRAS }

\newcommand {\be} {\begin{equation}}
\newcommand {\ee} {\end{equation}}
\newcommand{\beqa}{\begin{eqnarray}}
\newcommand{\eqa}{\end{eqnarray}}
\newcommand{\bea}{\begin{eqnarray}}
\newcommand{\eea}{\end{eqnarray}}
\newcommand {\bc} {\begin{center}}
\newcommand {\ec} {\end{center}}

\newcommand{\Msun}{\mbox{$\rm M_{\odot}$}}

\def\spose#1{\hbox to 0pt{#1\hss}}
\def\lta{\mathrel{\spose{\lower 3pt\hbox{$\mathchar"218$}}
        \raise 2.0pt\hbox{$\mathchar"13C$}}}
\def\gta{\mathrel{\spose{\lower 3pt\hbox{$\mathchar"218$}}
        \raise 2.0pt\hbox{$\mathchar"13E$}}}

\def\msol{\rm M_\odot}

\def\msol{\rm M_\odot}

\begin{document}

\begin{frontmatter}



\title{ADAFs, accretion discs and outbursts in compact binaries}


\author[label1,label2]{Jean-Pierre Lasota}

\address[label1]{Institut d'Astrophysique de Paris, UMR 7095 CNRS, UPMC Univ Paris 06, 98bis Bd Arago, 75014 Paris, France}
\address[label2]{Astronomical Observatory, Jagiellonian University, ul. Orla 171, 30-244 Krak\'ow, Poland}
\ead{lasota@iap.fr}

\begin{abstract}
I discuss the status of the Soft X-ray Transient model. First, I discuss
and then compare with observations the assumption that the
geometrically thin disc evaporates into an ADAF. Second, I address the
problems created by the recent determinations of the distance to
SS~Cyg, according to which the Disc Instability Model does not apply
to this famous dwarf-nova, thus casting doubt on the application
of this model to any system at all.

\end{abstract}


\end{frontmatter}

\section{Introduction}
\label{sect:intro}

As shown by \citet{dubus01} the disc instability model \citep[DIM,
see][for a review]{NAR} of Soft X-ray Transients (SXTs) must take
into account disc truncation and X-ray irradiation. Irradiation is
clearly observed in persistent systems (i.e. in Neutrons-Star (NS)
SXTs since all Black-Hole (BH) SXTs, are transient) and though it
had often been ignored or incorrectly described \citep[see][for the
correct description]{dubetal-99} its effects on the disc structure
in general and outbursts in particular are now universally accepted
and non-controversial. Disc truncation has a different status.
Although it is usually accepted that in low accretion-rate states
discs are truncated \citep[the famous Fig. 1 in][is now used as the
standard representation of the disc structure]{esin}, the reasons
usually given are spectral or timing-based. The fact that without
truncation it is impossible to get outburst cycles resembling the
observed ones \citep[as shown e.g. by][]{menou_trunc,dubus01} is
totally ignored even in the best reviews of the subject
\citep[e.g.][]{donetal}.

The reason for truncation is twofold. First, it is needed to get
very long recurrence times without playing games with the viscosity
parameter $\alpha$. Second, for non-truncated discs the DIM predicts
ridiculously low accretion rates onto the compact object, rates that
are clearly contradicted by X-ray observations of quiescent SXTs
\citep[see e.g.][]{lasota96,lny,ham-q}. This reason can be
easily bypassed. $\alpha$ is too simple a parameter to
describe the wealth of phenomena in accretion discs, and in the
standard DIM two values of this parameter (different
for cold and hot discs) must be used anyway.

The second reason, however, is impossible to evade (though not for
lack of trying). The idea behind the attempts at evading the low
accretion-rate restriction is very simple: if the X-rays in
quiescence are not accretion-generated then there are no constraints
on the accretion rate. The origin of quiescent X-rays had been
attributed both to the companion \citep{bild_rutl} and the disc
(except for the hot spot not much is left anyway) when they could
not be ascribed to the compact object, i.e. in the case of BH SXTs.
It was quickly demonstrated that the companion star cannot be the
source of X-rays in quiescence \citep{lasota00,ngm2}. The disc
origin of X-rays \citep{nay_sven} generated less interest.
Presumably the main reason was that a similar (in fact identical)
problem is encountered when applying the DIM to dwarf nova
outbursts. There too the X-ray luminosity is well above the level
predicted by the model, but in eclipsing systems the X-ray source is
clearly seen to be located near the white-dwarf. The size of this
source is no bigger than that of the white dwarf. Clearly it is not
the accretion disc. Of course one could argue that (truncated) discs
around black holes are totally different from CV discs, and in
addition to playing with $\alpha$ one could make them radiate in
X-rays; but such a solution is not very compelling. Astrophysicists,
in contrast to cosmologists and (post-modern) theoretical
physicists, usually adopt the ontological parsimony of William of
Ockham.
\begin{figure*}
\center
\includegraphics[width=1.9\columnwidth]{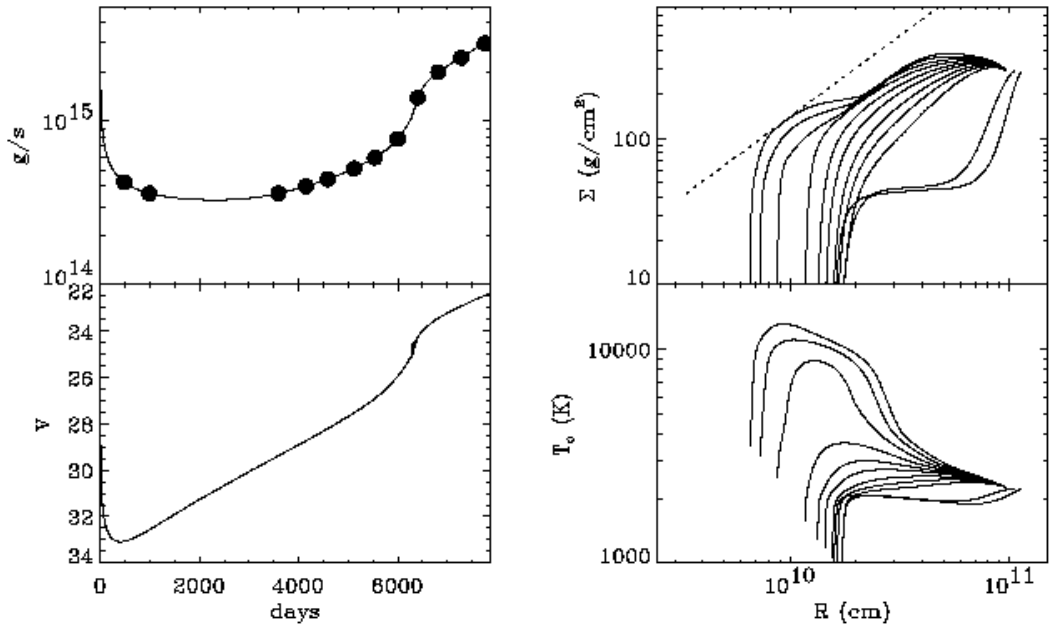}
\caption{Quiescence for one of the models discussed
    in \citet{dubus01}. The four panels show: the accretion rate $\dot{M}_{\rm in}$
    onto the black hole ($M_{\rm BH}=7 \Msun$, the V magnitude and the surface density
    and temperature profiles. The disc is truncated by ``evaporation".
    The mass transfer from the secondary is slow enough
    for matter to diffuse down the disc, gradually increasing
    $\dot{M}_{\rm in}$ and decreasing $R_{\rm in}$.  The outburst
    is triggered at $R\approx 10^{10}$~cm when $\Sigma$ reaches
    $\Sigma_{\rm max}$ (dotted line). The lower densities at the
    beginning of the quiescent state are due to the
    irradiation-controlled outburst decay. During the quiescence most
    of the surface-density profile is roughly parallel to the critical
    one $\sim R^{1.11}$.}
\label{fig:qui}
\end{figure*}

Although disc truncation is controversial (and its physics not
totally understood) it is almost universally agreed that the
mechanism responsible for SXT outbursts is described basically by
the same DIM that is used to describe dwarf-nova outbursts. Only the
effects of X-ray irradiation of the outer disc have to be added to
this model to make it work for SXTs. Truncation is the common
feature of both variants of the model. For example \citet{sscyg1}
showed that the multi-wavelength properties of the outburst cycle of
the dwarf-nova SS Cyg are best reproduced by the DIM with a
truncated inner disc. However, according to recent distance
determinations the DIM should not apply to SS Cyg because it is not
a dwarf nova. I will discuss this problem
\citep[following][]{sscyg2} in Sect. \ref{sect:sscyg}, after
addressing the question of the ``faint black-hole" paradigm.

\section{Are accreting black holes fainter than accreting neutron stars?}
\label{sect:faint}

The existence of truncated quiescent discs leads to several
problems. First, that of the truncation mechanism. In the case of
white dwarfs and neutron stars the discs can be truncated by the
magnetic fields of the accreting objects. Systems in which this is
the case are well known: polars (which truncate discs to
nonexistence), intermediate polars and of course accreting pulsars.
In the case of systems which have low states (and of discs in the
high state) no direct evidence is available for the moment but the
circumstantial evidence is pretty strong for VY Scl stars
\citep{hl_vy1,hl_vy2} and Aql~X-1 \citep{aqlx1}. However,
astrophysical black holes being devoid of their own magnetic fields,
disc truncation must be attributed to a different mechanism (which
could also work for some white dwarf and neutron star systems). This
mechanism, usually called ``disc evaporation", still escapes our
understanding despite some interesting contributions
\citep[e.g.][]{meyer_evap}.

The second problem concerns the form of the accretion flow beneath
the truncation radius. The simplest and oldest model is that of ADAF
\citep{nmy,lny}. But there are many other possibilities \citep[][but
see \cite{adiosadios}]{adios} and there is now evidence that
quiescent BH SXTs produce jets \citep{gallo_620} so it is possible
that a substantial part of the accretion flow does not cross the
horizon but is ejected from the system (possible but not necessary
if jets are produced at the cost of the black hole's rotational
energy as in the Blanford-Znajek mechanism). Mixed models (ADAF +
jet) were therefore suggested \citep[see e.g.][]{yuan_narayan}. One
should remember here that even in the case of a ``pure" ADAF (no
outflow) the thermal energy advected into the black hole can be (in
principle) negligible \citep[][and Fig. 2 in
\cite{chen_glob}]{bp-98}.

The ADAF solution applied to both accreting black holes and neutron
stars has one immediate consequence: neutron stars should be
brighter because all the thermal energy that was not radiated away
from the accretion flow will have to be emitted from the stellar
surface. However, one should \textsl{not} deduce from this that any
quiescent black-hole binary system will be fainter than any
neutron-star binary in a similar state of activity. The ``faint
black-hole paradigm" is supposed to apply only to compact bodies
accreting at the \textsl{same} rate. In \citet{CRP} I recalled the
brief and unfinished story of testing the ``faint black-hole
paradigm". Here I will discuss in more detail the method
\citep{somelike,menouetal} of attempting to make sure that the BH
and NS SXTs under comparison have comparable accretion rates.

The idea is to plot the quiescent luminosity as a function of the
orbital period (Fig. \ref{fig:faint}). Contrary to naive
expectations, this method does \textsl{not} assume that BH SXTs and
NS SXTs have the same transfer rates at a given orbital period. It
is based on the assumption that the truncation radius where the
transition from disc to ADAF occurs is roughly a constant fraction
of the circularization radius \citep{menouetal,faint,lasota00} \be
R_{\rm tr}=fR_{\rm circ}(M_1,q,P_{\rm orb}), \ \ \ f \lta 0.48,
\label{eq:retrunc} \ee $q$ being the mass ratio (secondary/primary),
$M_1$ the primary mass, and the value of 0.48 coming from the
requirement that the inner disc radius must be no smaller than the
impact distance of the mass-transfer stream.

The accretion-rate profiles in a quiescent disc are roughly parallel
to the critical one $\dot{M}_{\rm crit}\sim R^{2.68}$, but close to
the outer disc edge they flatten to match the mass-transfer rate(see
e.g. Fig. \ref{fig:qui}). Truncating the disc at radius given by Eq.
(\ref{eq:retrunc}) determines an accretion rate which is independent
of the mass-transfer rate. The resulting accretion rate will depend
on $M_1$ and $q$ but this would make the ratio between the NS and BH
SXT accretion rates differ by a factor 2 - 4 and not by more than an
order of magnitude, as observed. In fact on Fig. \ref{fig:qui} one
could attribute to the luminosities, as a function of period, a
slope of roughly 1.7 -- 1.8, as expected from the slope of the
$\dot{M}_{\rm crit}(R)$ relation \citep[see][]{ham-q}.

One can compare the assumptions described above with models of SXT
outburst cycles by \citet{dubus01}. Figure \ref{fig:qui} shows the
quiescent phase of a BH SXT outburst cycle. The right panels
represent a dozen surface density $\Sigma$ (upper panel) and central
temperature $T_c$ (lower panel) profiles of a quiescent accretion
disc around a black hole. Dots on the ``light-curve" (upper-left
panel) show where the ``snapshots" were taken. Except for the first
two (just after the end of the outburst) and the last two (just
before the next outburst) profiles, the surface-density profiles are
roughly parallel to the critical-density lines (in fact they are
slightly shallower). During $\sim 20$ years the inner radius moves
in by a factor $\sim 3$ but until the last $\sim 3$ years the
accretion rate does not vary  by more than a factor of 2. This
reflects the ``self-similar" way the disc fills up. \citet{dubus01}
discuss the dependence of the outburst cycle properties on the mass
of the compact object. Although, as mentioned above, there is some
$M_1$ dependence in the critical values of density (accretion rate),
the strongest mass-dependence enters into the model trough the
formula describing the disc truncation, i.e. the disc
``evaporation", because $\dot M_{\rm evap} \sim M_1^3$. As a result,
in quiescence, the disc around a lower-mass compact object is
truncated at smaller radii, which reduces the inner accretion rate.
This effect would therefore reduce the ratio of neutron-star to
black-hole quiescent luminosities. Of course one should keep in mind
that the evaporation prescription used here is not based on any
physical mechanism but other physically motivated approaches are not
easily used in full irradiated-disc outburst calculations \citep[as
discussed in][]{dubus01}.

One can conclude that testing the faint black hole paradigm by
comparing quiescent NS and BH SXTs at similar orbital periods is a
reasonable procedure (and in any case the only one that makes
sense).

\subsection{The case of 1H~1905+000}

Recently \citet{jonker1,jonker2} found that the neutron-star soft
X–ray transient 1H~1905+000 could be the spoilsport long-awaited by
the anti-ADAF crowd. Its quiescent X-ray luminosity is at most $\sim
1 \times 10^{30}$ erg s$^{-1}$. \citet{jonker2} notice that ``this
luminosity limit is lower than the luminosity of A0620-00, the
weakest black hole soft X-ray transient in quiescence reported so
far", which is a true statement. However, the conclusion that ``the
claim that there is evidence for the presence of a black hole event
horizon on the basis of a lower quiescent luminosity for black holes
than for neutron stars is unproven" is not. After the publication of
the first upper limit in \citet{jonker1} I addressed this point in
\citet{CRP}, explaining in detail why the faintness of the neutron
star in 1H~1905+000 does not invalidate the fainter black hole
paradigm. However, since my discussion was overlooked by
\citet[][but see \cite{cornelisse}]{jonker2}, I think it will be
useful to repeat the argument in slightly different form, using
formulae freshly calculated for helium discs in \citet{he}.

\begin{figure*}
\center
{\includegraphics[width=1.5\columnwidth]{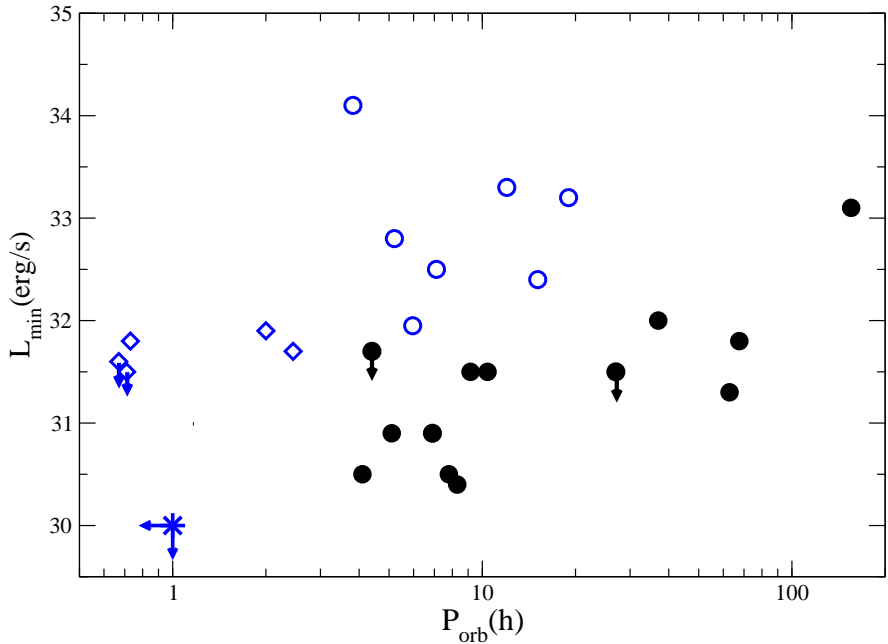}}
\caption{Quiescent luminosities of black holes (filled circles) and
neutron star
 (open circles and open diamonds) soft X-ray transients. Diamonds
 correspond to accreting millisecond pulsars. The star represents
 the system 1H~1905+000 whose orbital period is unknown but all the available
 evidence suggests it is a UCXBs (see the
 text). This figure shows that in quiescent
 transient LMXBs, \textsl{at a given orbital
 period}, neutron stars are brighter than black holes.
 Data for black holes from Garcia (private communication),
 for neutron stars from \citet{Campana}.}
\label{fig:faint}
\end{figure*}

I will use Fig. \ref{fig:faint} which the same as Fig. 1 in
\citet{CRP} but with a modified position of 1H~1905+000. As before,
the orbital period of 1H~1905+000 is unknown but it is certainly
very short \citep{jonker2}. I have therefore tentatively assumed a
period of $1$ h, but even a longer period would not contradict the
claim that black-hole systems are fainter than those harbouring
neutron stars (no black-hole low-mass X-ray binary has been observed
with an orbital period $\lta 4$ hr).

Even if the actual quiescent X-ray luminosity of 1H~1905+000 were
much lower than the \textsl{Chandra} upper limit it would not
necessarily be a problem for the black-hole faintness paradigm. This
is because 1H~1905+000 is a rather unusual binary. The faintness of
the secondary implies an ultra-compact X-ray binary (UCXB), in which
case the neutron star companion would be a low-mass helium or
carbon-oxygen white dwarf \citep{ucb}. When transient \citep[very
short-period systems are rather persistent][]{deloy}, such compact
binaries exhibit short ($\gta$10 -- $\gta$100 days), exponentially
decaying outbursts, as expected from small, X-ray irradiated
accretion discs \citep{dubus01}. In all these very compact transient
systems the neutron star is a millisecond pulsar (MSP). Both their
outburst (usually a few $\%$ of the Eddington luminosity) and
quiescent X-ray luminosities ($< 10^{32}$~erg s$^{-1}$) are lower
than those observed in longer period SXTs \citep{Campana}. This
might seem to be similar to 1H~1905+000 whose outburst luminosity
was $\sim 4\times 10^{36}$ erg s$^{-1}$. However, the outburst
behaviour of this system is totally different from that observed in
other short-period binaries and UCXBs. Instead of short outbursts
1H~1905+000 exhibited one $\gta$ 10 year long outburst that ended in
the late 1980s or early 1990s. Since then it has been quiet.
It is not clear why 1H~1905+000 is so different. During 11 years,
say, it accreted $\sim 8 \times 10^{24}$ g. This is a lot, but a
helium accretion disc can contain as much as
\begin{eqnarray}
M_{\rm D, max}\approx  3.5 && \times 10^{25}\left(\frac{\alpha_{\rm
cold}}{0.01}\right)^{-0.83}\nonumber\\
&&\times\left( \frac{M_{\rm ns}} {\rm 1.4
M_\odot} \right)^{0.67} \left( \frac{P_{\rm orb}}{1\rm h}
\right)^{2.13}\ {\rm g},
\label{eq:diskmass}
\end{eqnarray}
where $\alpha_{\rm cold}$ is the cold-disc viscosity parameter
and $P_{\rm orb}$ the orbital period.
I used the helium-disc critical surface density from \citet{he}.
For the disc radius I took:
\be
\frac{r_{\rm D}(\rm max)}{a}=\frac{0.60}{1+q},
\label{eq:rd_bp}
\ee
\citep[valid for $0.03 < q< 1$,][]{bp-77}, where
\be
a = 2.28 \times 10^{9}M_1^{1/3}(1 + q)^{1/3}P_{\rm min}^{2/3}\,\rm cm
\label{eq:separ}
\ee
is the binary separation; $P_{\rm min}$ being the orbital
period in minutes. In Eq. (\ref{eq:diskmass}) $r_{\rm
D}=0.6\,a$

The maximum outburst luminosity for an irradiated helium disc around a 1.4
$\msol$ neutron star can be estimated as
\begin{equation}
L_{\rm max} \simeq 3.5 \times 10^{37} \left( \frac{P_{\rm orb}}{1
\rm h}\right)^{1.67}\rm erg\ s^{-1},
\label{lmax}
\end{equation}
\citep{he}.
Therefore 1H~1905+000 could in principle be a ``normal",
short-duration X-ray transient source, but it isn't. Maybe its long
"outburst" was due to irradiations of the secondary. If its period
is $\sim 20$ min it could be marginally stable with respect to the
thermal-viscous instability in an irradiated helium (or
carbon-oxygen) accretion disc.

\begin{figure}
\includegraphics[width=\columnwidth, angle=0]{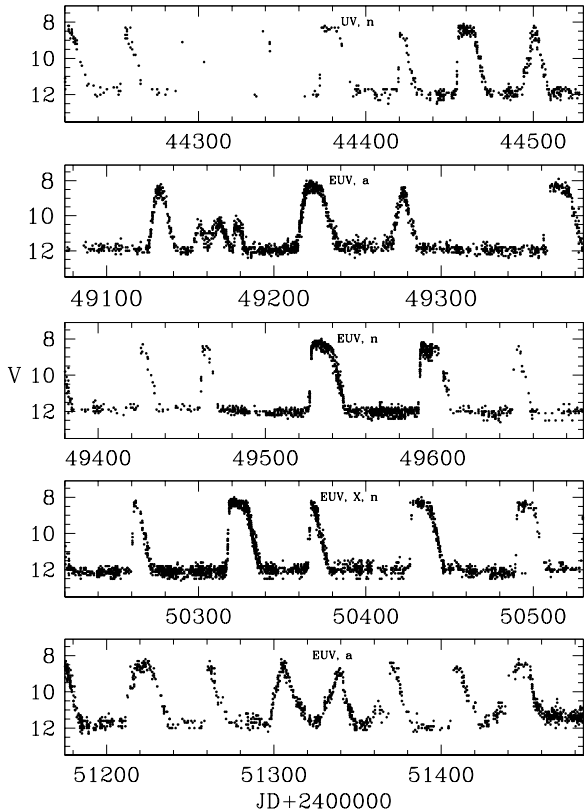}
\caption{\label{f-sscvis} Examples of the visual light curve of
SS\,Cyg. Outbursts called normal are marked with ``n" and so-called
anomalous outbursts are designated by ``a". SS\,Cyg is also a UV,
EUV, and X-ray emitter. The data are from the AFOEV, the figure from
\citet{sscyg1}.}
\end{figure}
Another possibility is that the companion star stopped transferring
mass. In such a case, after a short time, the neutron star would
cease accreting at all. It is perhaps worth repeating that in such a
case the \textsl{accretion} luminosity of the neutron star will
certainly be lower than the luminosity of any accreting black hole.
Then the absence of X-rays is a problem for \citet{brown_bild} but
not for me or other ADAF-advocates. Such a scenario is quite
realistic. After all there exist CVs which do exactly that: some
VY~Scl stars stop transferring mass for years after being for years
among the brightest CVs. We don't know why but they do. These
strangely behaving companion stars are K-M dwarfs (close to being
fully convective) while the secondary of 1H~1905+000 is most
probably a helium or C/O star, but this should not prevent them from
arresting mass transfer.

The considerations above assume that the binary systems in question
have a \textsl{bona fide} structure of a LMXB: matter flowing from
the $L_1$ point forms a disc through which it is accreted onto the
compact object. However, the form of mass-transfer in systems with
such very low mass ratios has not been studied and only some general
properties of such systems can be conjectured~(Dubus, private
communication). For values of $q\lta 0.02$ the circularization
radius becomes larger than the estimates of the outer radius
Eq. (\ref{eq:rd_bp}). Most probably matter streaming in from the
companion circularizes onto unstable orbits. For $q\approx 0.02$,
matter is added at $R_{\rm circ}$ onto orbits that can become
eccentric due to the 3:1 resonance. For $q\approx 0.005$ the
circularization radius approaches the 2:1 Lindblad resonance. This
might efficiently prevent mass being transferred onto the compact
object.

\textsl{Nota bene}, equivalent systems with a black-hole instead of
a neutron star would have a minuscule mass ratio $< 0.01$ ($M_{\rm
bh}> 4 \msol$). It is probably not a coincidence that there are no
observed black-hole counterparts of neutron-star X-ray binary
systems at orbital periods shorter than 2 hours (it is not certain
if this could explain the lack of observed BH LMXBs below 4 hours,
though). Evolutionary models suggest that such systems should exist
\citep{ljova}. If they do, they are not your normal LMXBs.

Although I agree with Danny Steeghs (private communication) that
``there is always something [to be said] for at least trying to kick
paradigms and see if they fall over", I think the 1H 1905+000 kick
is too weak even to shake the faint black hole paradigm. But
observers should of course keep trying.

\section{Does the thermal-viscous instability model describe dwarf-nova outbursts?}
\label{sect:sscyg}

SS\,Cyg is the brightest and best observed dwarf nova in the sky
(see Fig \ref{f-sscvis}). It therefore came  as a shock when its
distance (reputed until then to be $\sim 100$\,pc) was announced to be
$166\pm12$\,pc after the determination of the HST/FGS parallax of
this system \citep{harrisonetal99-1}. It was immediately noticed
that at such a distance the mass-transfer rate would be well above
the critical one and as a consequence, according to the DIM, SS\,Cyg
should not be a dwarf-nova contrary to observational evidence.
\citet{gan_schr} analyzed the consequences of this drastic distance
revision and came at the conclusion that if we want to avoid a
substantial revision of the model we must either assume an important
enhancement of the mass transfer during the outburst (which would in
fact be a substantial revision of the model anyway) or bring back
the distance close to the previous value. This latter choice was
probably that of the majority, at least of those who noticed the
problem.
\begin{figure*}
\center
{\includegraphics[width=1.5\columnwidth,angle=-90]{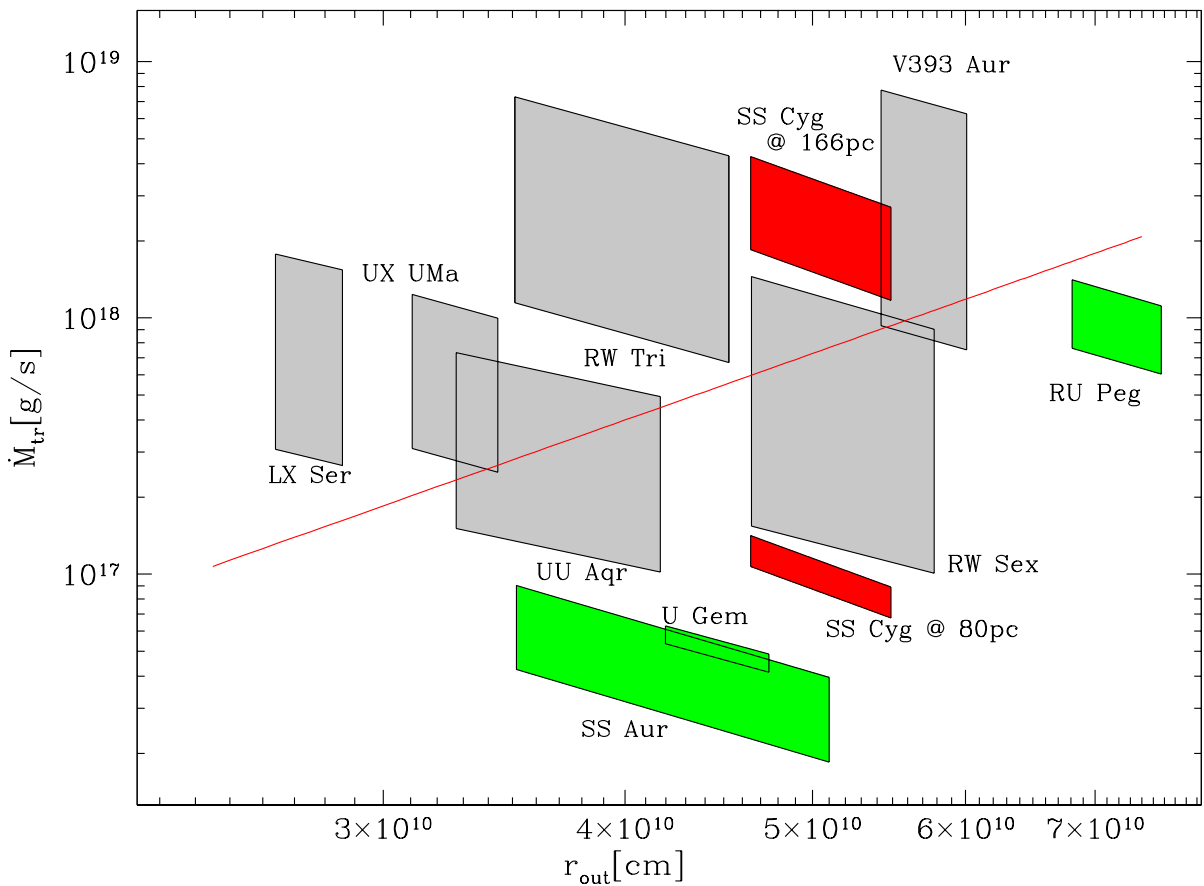}}
\caption{\label{fig_mtr_all} The mean mass transfer rate of SS\,Cyg
and of six well known nova-like CVs as a function of the outer
radius of the disc during outburst. As the binary parameter of
SS\,Cyg the values (and uncertainties) derived by
\citet{bitneretal07-1} were used. Both, in order to make the plot
easier to read and because the broad ranges of possible parameters
do not represent well-determined values with certain errors, we use
shaded boxes instead of error bars. The solid line represents the
critical mass transfer rate. According to the DIM, this line should
separate dwarf novae and nova-like systems. On the other hand SS\,
Cyg should be stationary, being brighter than (or as bright as)
nova-like systems \citep[for details see][]{sscyg2}.}
\end{figure*}
However, recently \citet{bitneretal07-1} re-determined the
parameters of SS\,Cyg and arrived at the conclusion that they are
consistent with a distance $\gta$ 140 -- 166 pc supporting the
parallax measurement. This independent confirmation of the large
distance to SS Cyg triggered a new re-evaluation of the validity of
the DIM \citep{sscyg2}.

First we confirmed that at a distance $\gta$ 140pc the mass-transfer
rate of SS Cyg is in contradiction with the DIM. However, we also
pointed out that the problem is more serious than the failure of the
DIM. In fact we know from observations of disc accretion in CVs that
below a certain mass-transfer rate their discs go into outburst,
producing dwarf novae. At higher mass transfer rates, the disc is
stationary and the corresponding class of CVs form the nova-like
systems. In agreement with this picture, the mean absolute
magnitudes of dwarf novae have been found to be lower than those of
nova-like systems \citep[see][Fig.\,9.8]{warner95-1}. To check
whether this agreement remains valid for a distance to SS\,Cyg of
$166\pm12$\,pc, we compared the mean mass transfer rate derived for
SS\,Cyg with that obtained for a set of well-observed nova-like
systems and three dwarf novae with measured HST/FGS parallax
\citep[see Tables 2 and 3 in][]{sscyg2}.

Fig\,\ref{fig_mtr_all} \citep[inspired by Fig 1 in][who was the
first to attempt such a test]{smak83-1} shows the derived mean mass
transfer rates as a function of the outer radius of the disc during
outburst. To avoid our results depending on uncertainties in the
system parameter derived from observation, we used rather broad
ranges of parameters. The straight line represents the critical mass
transfer rate for $_1=1\Msun$. Obviously, at a distance of $166$\,pc
the mean mass transfer rate of SS\,Cyg is above this limit, as
mentioned above. The other three dwarf novae are below the dividing
line and the nova-like stars have mass transfer rates higher than
(or similar to) the critical rate. The striking point of
Fig.\,\ref{fig_mtr_all} is the fact that the mean mass transfer rate
of SS\,Cyg is higher than (or as high) as those derived for
nova-like systems with similar system parameters. Therefore if
SS\,Cyg is indeed $d\gta140$\,pc away, the difference between
nova-like systems and the dwarf nova SS\,Cyg cannot be attributed to
the mean mass transfer rate. This conclusion obviously contradicts
the generally accepted picture of accretion discs in CVs, not only
the DIM.

Clearly, one could argue that the distances to the nova-like systems
are systematically too small. However, the distance to RW\,Tri is
based on a HST/FGS parallax and for the other systems we used very
large upper limits for the distance. Hence there is no easy way out
of the problem. A non-easy way is to assume that SS\,Cyg outbursts
are triggered by mass transfer enhancements. These enhancements
would have to be so serious that this would be equivalent to getting
back to the old universally rejected mass-transfer instability
model. This in itself would not be a problem (at least not for me).
The problem would be to explain why SS\,Cyg is does this while its
neighbours (in period and spectral type) do not. What is so special
about SS Cyg? Since this is a place where I can express my personal
feelings I will admit that I believe that finally it will be
established that the true distance is $\sim 100$\,pc.

In the case of SXTs there was a similar problem with GRO J1655-40.
At first it was thought to transfer mass at a rate too high for it
to be transient. Various attempts to circumvent this problem were
made but it was the revision of the parameters of this binary by
\citet{beer_pods} that solved the problem. Now we hear that
GRO~J1655-40 is much closer than previously thought  and this would
reduce the mass-transfer even more. But, paradoxically, if we were
one were to accept the distance proposed by \citet{mirabel}, the
secondary in GRO~J1655-40 would not be filling its Roche-lobe and in
this case too we would have to invoke a mass-transfer instability to
explain the outbursts.\footnote{It is rather unlikely that the
secondary in GRO~J1655-40 is not filling its Roche-lobe but the
mass-transfer enhancement still might contribute to the outburst
properties \citep[see][]{1650esin}.}

\section{Conclusion}

Of the two pillars of the SXT--outburst model, the ADAF one is
making quite a good job at resisting attempts to topple it. The
other one, the DIM, is in danger of falling if the larger distance
to SS Cyg is to be taken seriously. If the DIM were to be completed
by a contribution from enhanced mass-transfer from the secondary I
would be the last to be surprised \citep[see
e.g.][]{lasota96a,lasota97}.

\section*{Acknowledgements}

I am immensely grateful to my friend Marek Abramowicz for the
perfect organization of this highly successful and enjoyable
conference. This research used the Association Fran\c{c}aise des
Observateurs d'\'Etoiles Variables (AFOEV) database, operated at
CDS, Strasbourg (France).

\bibliographystyle{elsart-harv}


\end{document}